# Impact of Size Effect on Graphene Nanoribbon Transport

Yinxiao Yang and Raghunath Murali

*Abstract*: Graphene has shown impressive properties for nanoelectronics applications including a high mobility and a width-dependent bandgap. Use of graphene in nanoelectronics would most likey be in the form of graphene nanoribbons (GNRs) where the ribbon width is expected to be less than 20 nm. Many theoretical projections have been made on the impact of edge-scattering on carrier transport in GNRs – most studies point to a degradation of mobility (of GNRs) as well as the on/off ratio (of GNR FETs). This study provides the first clear experimental evidence of the onset of size-effect in patterned GNRs; it is shown that for W<60 nm, carrier mobility in GNRs is limited by edge-scattering.

## 1 INTRODUCTION

Graphene has been found to have a variety of interesting and superior properties such as high mobility[1], high thermal conductivity[2], width-dependent bandgap[3], and resistivity better than equivalent Cu wires[4]. There have been a number of efforts to build transistors out of graphene, for both RF and digital switching. Graphene on a $SiO_2$ substrate[5] can have a mobility as high as 40,000 $cm^2$/V-s whereas suspended graphene mobility has been measured[1] to be more than 200,000 $cm^2$/V-s.

It is well-known that as the dimensions of a material are scaled, significant changes occur in carrier transport: carrier quantization alters the charge distribution and edge/boundary scattering degrades mobility. A decrease in mobility of graphene nanoribbons (GNRs) is predicted for line-widths below 10 nm[6] where a significant bandgap opens up – the mobility is predicted to exponentially decrease as the bandgap becomes larger. In realistic GNRs, either lithographically patterned or chemically produced, a certain amount of line-edge roughness (LER) is expected. A number of theoretical predictions for graphene have been made to quantify the impact of LER-induced scattering on carrier transport. Metallic armchair and zigzag GNRs with edge roughness of just a few atoms exhibit significant conductance fluctuations[7]; in addition, even for metallic zigzag ribbons, a transport gap opens up for small widths. Edge disorder effectively wipes out any distinction between zigzag and armchair GNRs[8]. Simulations[9] show that for a moderately disordered armchair-GNR FET with W~5 nm, device on-off ratio is degraded by 10X from the ideal case of a perfect edge; even for an almost near perfect armchair GNR, a nearly 3X on-off ratio degradation is observed at this line-width. In another study, various scattering mechanisms and their respective limits on carrier transport are modeled[10] and it is predicted that the onset of LER-induced degradation of mobility occurs for line-widths less than 5 nm.

There have been a number of measurements of graphene mobility – most of these have been on wide ribbons [1, 11, 12]. Most narrow-GNR characterization [3, 13-15] has been in the context of bandgap opening. There has been little experimental work on characterizing the impact of line-width scaling on GNR transport. In chemically derived GNRs[14], it has been observed before that a smaller width leads to a smaller mobility although a relationship has not been extracted. In this work, GNRs of widths between 16 nm and 1 μm are fabricated, and their mobility is extracted to correlate the impact of lateral scaling on carrier transport.

## II EXPERIMENT

Few layer graphene (1-8 layers) is used as the starting material. Monolayer and bilayer graphene is identified by confocal micro-Raman imaging. For three or more layers, AFM scanning and optical imaging is used to estimate the number of graphene layers. A first lithography step defines metal contacts (Ti/Au stack) while a second lithography step defines nanometer-wide channels with width (W) in the range 16nm<W<1000nm, and a length of 0.4 μm and 0.7 μm. A low-power oxygen plasma etch is used to transfer resist patterns into the channel. Four-point probe measurements were performed with a standard lock-in amplifier and back-gated measurements with a semiconductor parameter analyzer. The main metric used in this work to compare across dimensions is mobility; since mobility changes with carrier concentration, it is important to measure mobility at the same carrier density across different GNRs to ensure a fair comparison. The back-gate thickness is 300 nm, and fringe-field effects, simulated using COMSOL, are accounted for in extracting mobility.

Fig. 1 shows a representative device studied in this work; there are 10 GNRs, all with identical widths, between each set of electrodes. The mobility of GNRs is extracted for 6 samples, each with 3-6 devices of the type shown in Fig. 1; a total of 21 devices were studied. The inset of fig. 1 shows gate characteristics, $I_{ds}$-$V_{gs}$, for devices of varying width, from which mobility is evaluated. Four-point probe measurements reveal that contact resistance is only a small portion (<2%) of the overall resistance – this is expected since contacts are made to large graphene areas. Mobility (μ) is extracted using the carrier density method: $\mu_n = 1/(n\rho e)$ where $n=5\times10^{12}$ $cm^{-2}$, ρ is resistivity, and e is electronic charge. An alternative method, via trans-conductance, is calculated by $\mu_{gm}=g_m/(C_{ox} \cdot W/L \cdot V_{ds})$, whereby $g_m$ is evaluated at the equivalent $n=5\times10^{12}$ $cm^{-2}$ carrier density. The trans-conductance method is less accurate than the carrier density method due to non-linear IV behavior, a result of short-range scattering from ripples and point defects [16]. Hence, we strictly utilize the carrier density method for mobility extraction in our study.

Carrier mobility is extracted at various stages during the patterning of graphene: (1) after contact metallization ($\mu_1$): the mobility extracted here is that of a microns-wide region rather than that of GNRs; measurements at this stage are also used to extract the impurity density based on Dirac point shift[12]; (2) after resist definition ($\mu_2$): HSQ resist is patterned using e-beam lithography, and developed in a TMAH developer, leaving behind fine HSQ patterns on the graphene flake; mobility extracted during this step provides insight into how the overlaying dielectric layer affects carrier transport;

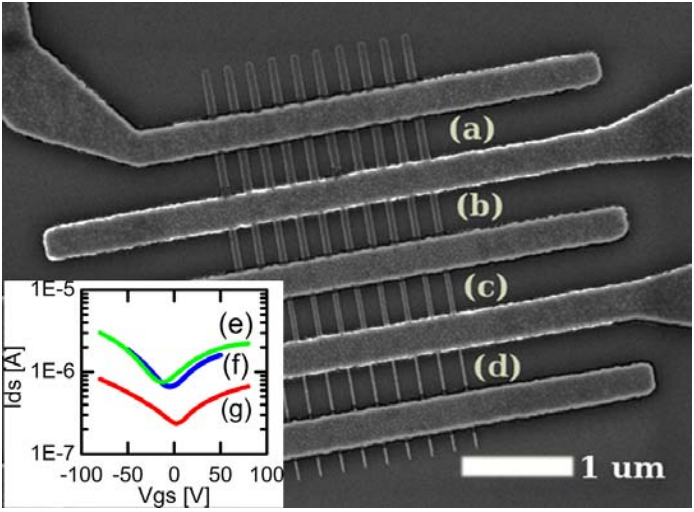

Figure 1: SEM image of a device with five electrodes; there is a set of 10 parallel GNRs between an electrode pair. The GNRs (below HSQ lines) are 68nm (a), 58nm (b), 42nm (c), and 20nm (d) wide and have a length of 400nm. The inset shows gate characteristics, $I_{ds}$-$V_{gs}$, for devices of varying width – (e) 50nm, (f) 45nm, and (g) 25nm.

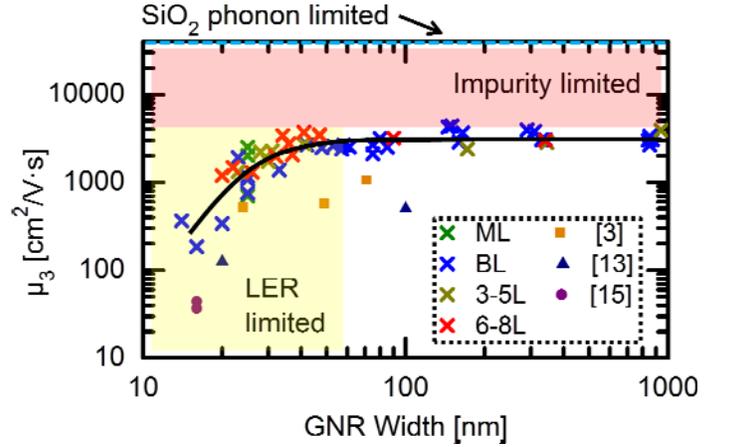

Figure 2: Total mobility versus line-width: 54 data points from the current work categorized by layer thickness: monolayer (ML), bilayer (BL), 3-5 layers, and 6-8 layers. Also plotted are 7 data points from previously published work [3, 13, 15]. For W<60nm, the size-effect is seen to degrade GNR mobility (yellow region); for W>60nm, mobility is limited by impurity scattering (pink region).

also, any Dirac point shift at this stage compared to step-1 can be attributed to charge induced by the HSQ layer; (3) after plasma etch ($\mu_3$): the HSQ resist pattern is etched into the graphene flake; mobility extracted at this step is that of patterned graphene ribbons. The LER has been characterized using an SEM and is between 1-3 nm. The oxygen-plasma etch step has been found to have little effect on the LER as found from pre-etch and post-etch SEM imaging.

### III ANALYSIS

It is found that an HSQ coating does not degrade carrier mobility – sometimes it actually improves mobility; this is consistent with some previous findings where PMMA and other layers were found to have little impact on graphene mobility[17]. However, $\mu_3$ is seen to be always smaller than either $\mu_1$ or $\mu_2$ – this indicates that conversion of graphene to GNRs is likely to be the main reason for mobility degradation.

Impurity density ($n_{imp}$) has a strong relationship with device mobility; the origin of impurity scattering remains unclear but evidence points toward the impurities in the $SiO_2$ substrate. Impurity-limited mobility ($\mu_{impurity}$) is estimated using previously published techniques [12]. Devices fabricated in this work have an impurity density in the range $1.1 \times 10^{11}$ to $2.2 \times 10^{12}$ cm$^{-2}$. This corresponds to an impurity-limited mobility range of 48,000 cm$^2$/V-s to 2,200 cm$^2$/V-s. To portray the size-effect discussed in this work, it is better to compare devices with a similar impurity density. Thus, devices with $1 \times 10^{12}$ cm$^{-2}$ < $n_{imp}$ < $2 \times 10^{12}$ cm$^{-2}$ are chosen to plot total mobility versus line-width, Fig. 2. Also shown for reference are a few previously published results with W<100 nm [3, 13, 15]. The mobility for W>100 nm is seen to be limited by impurity density scattering; at these line-widths, if impurity density was kept to below $1 \times 10^{11}$ cm$^{-2}$, the mobility would then be limited by $SiO_2$ phonon scattering. For W<60nm, there is a trend of decreasing mobility with decreasing line-width – this is a manifestation of the size-effect. It can be seen that mobility of a GNR in Fig. 2 decreases from more than 3000 cm$^2$/V-s for W~100 nm to less than 200 cm$^2$/V-s for W<20 nm. To extract the size-dependent component of mobility, we take the difference between $\mu_2$ and $\mu_3$. Since the smallest line-width is 16 nm, effective mass can be assumed nearly the same [6] for the range of line-widths considered in this work, and the dominant mechanism for the size-effect is from edge-scattering.

After step-2, graphene is still in its 2D form, and thus the LER-limited mobility does not limit total mobility ($\mu_2$). After step-3, 2D graphene is converted to 1D graphene, and now the LER component of mobility ($\mu_{LER}$) becomes a limiting factor of the total mobility ($\mu_3$). An assumption made here is that step-3 (the plasma etch step) does not contribute to a degradation of mobility either by defect generation in graphene or by some other mechanism. This assumption has been experimentally verified by sequentially etching various devices – after the initial decrease, no further decrease in mobility is seen even for long periods of plasma etch. This is reasonable to expect since the graphene that remains after the etch is being protected by HSQ resist. Fig. 3 shows the LER-limited mobility plotted against width. A reciprocal relationship is seen between LER-mobility and line-width; the data fit is of the type $\mu_{LER} = A \cdot W^B$ where A is a constant and B=4.3. This value of B compares well with a previous model which predicted B=4.0 [10]. The result in Fig. 3 shows that LER-induced mobility degradation sets in at a width of around 60 nm; this is much wider than the sub-5 nm widths predicted earlier. The inset of fig. 3 shows the overall mean free path (MFP) versus line-width; MFP decreases from about 100 nm for W>100 nm to less than 10 nm for W< 20 nm.

Carrier mobility in graphene is limited by various scattering mechanisms[5, 10]: (i) intrinsic scattering that limits mobility to 200,000 cm$^2$/V-s, (ii) $SiO_2$ phonon scattering that limits mobility to 40,000 cm$^2$/V-s, (iii) impurity scattering, and (iv) LER scattering. Using Matthiessen's rule, the total mobility can be written as

$$\frac{1}{\mu} = \frac{1}{\mu_{lattice}} + \frac{1}{\mu_{SiO2}} + \frac{1}{\mu_{impurity}} + \frac{1}{\mu_{LER}} \quad (1)$$

A trend-line is obtained by using equation (1) with $\mu_{impurity}$ estimated to be 2,700 cm$^2$/V-s and $\mu_{LER}$ from the fit in fig. 3.

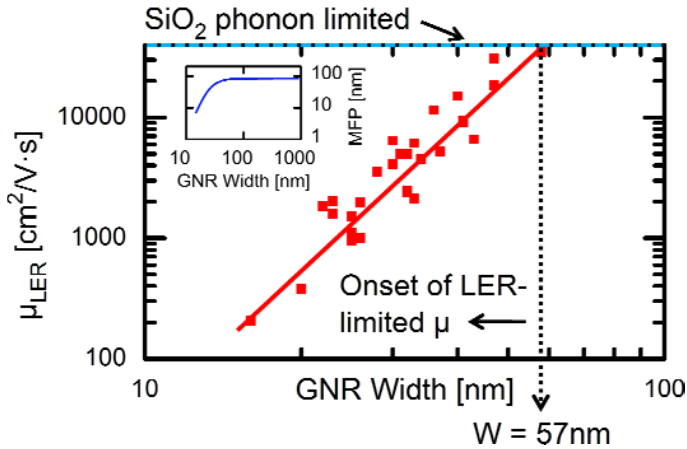

Figure 3: LER-limited mobility is seen to have a dependence of the type $W^{4.3}$. Below W=57nm (where the fitted line intersects with the $SiO_2$ phonon limited mobility value), the size-effect due to edge scattering leads to a degradation of mobility. The inset shows device mean free path (MFP) versus line-width.

This model is plotted in fig. 2 as a solid line and agrees well with the data obtained in this work. Mobility data from previous work are all smaller than the mobility found in this work. This could be due to a combination of worse impurity scattering, different starting material (HOPG versus Kish graphite), and worse LER occurring in previously fabricated devices. Extrapolating the trend-line in fig. 2 to the W~2 nm devices found in [14], the model predicts that mobility is less than 10 $cm^2$/V-s whereas the actual mobility found is 50-100 $cm^2$/V-s. This is clearly due to a difference in the way GNRs are created in [14]: by chemical methods rather than mechanical exfoliation and top-down lithography used in this work. The better mobility can be explained by the fact that chemical methods may show smoother edges; but even these smoother edges are not sufficient to obtain high-mobility GNRs. In addition, chemical methods of producing GNRs present challenges in nanoribbon placement. Thus, methods need to be found to combine top-down and bottom-up approaches to fabricating GNRs so that edge scattering is minimized.

## IV CONCLUSION

The impact of line-width scaling has been correlated to graphene nanoribbon transport. GNR mobility is found to be severely limited by LER-scattering for W<60 nm. While a mobility of more than 3000 $cm^2$/V-s can be achieved for W>60 nm, the value decreases to less than 200 $cm^2$/V-s for W<20 nm. This trend agrees well with the expected impact of LER-limited mobility though previous predictions about the onset of LER-limited mobility are widely off. The size-effect demonstrated in this work severely constrains the use of GNRs for nanoelectronics applications unless methods are found to produce high-quality GNRs.

## ACKNOWLEDGEMENT

The authors would like to acknowledge support from the Nanoelectronics Research Initiative (NRI) through the INDEX program.